\documentclass[sigconf,screen,authorversion,nonacm]{acmart}

\usepackage{fancyhdr}
\AtBeginDocument{%
    \addtolength{\footskip}{2.0\baselineskip}%
    \fancyfoot[L]{\textit{\textbf{Preprint --- Published in ACM BSCI'23. }}}%
}
\usepackage{balance}
\usepackage{graphicx}
\usepackage{xcolor}
\usepackage{tabularx}
\graphicspath{{fig/}}

\newcommand{\sig}[2]{\sigma_\mathit{#1}^\mathit{#2}}
\newcommand{\sln}{DHTee}
\newcommand{\rqst}{\mathit{rqst}}

\newcommand{\rpt}{\mathit{rpt}}
\newcommand{\natv}{\mathit{natv}}

\newcommand{\attrqst}{\texttt{AttRqst()}}
\newcommand{\attgen}{\texttt{AttGen()}}
\newcommand{\attconvrfy}{\texttt{AttConVrfy()}}
\newcommand{\attvrfy}{\texttt{AttVrfy()}}

\newcommand{\eqname}{Equation}
\newcommand{\para}[1]{\smallskip\noindent\textbf{#1.}}

\AtBeginDocument{%
  \providecommand\BibTeX{{%
    \normalfont B\kern-0.5em{\scshape i\kern-0.25em b}\kern-0.8em\TeX}}}

\begin{document}

\title{Decentralized Translator of Trust: Supporting Heterogeneous TEE for Critical Infrastructure Protection}
%\title{\sln{}: Decentralized Infrastructure for Heterogeneous TEEs}

\author{Rabimba	Karanjai}
\orcid{0000-0002-6705-6506}
\email{rkaranjai@uh.edu}
\affiliation{%
  \institution{University of Houston}
  \city{Houston}
  \country{US}
} 

\author{Rowan Collier}
%\orcid{0000-0002-7662-2119}
\email{rcollie9@kent.edu}
\affiliation{%
  \institution{Kent State University}
  \city{Kent}
  \country{US}
} 

\author{Zhimin Gao}
\email{mtion@msn.com}
\affiliation{%
  \institution{Auburn University at Montgomery}
  \city{Montgomery}
  \country{US}
}

\author{Lin Chen}
\email{chenlin198662@gmail.com }
\affiliation{%
  \institution{Texas Tech University}
  \city{Lubbock}
  \country{US}
}

\author{Xinxin Fan}
\email{xinxin@iotex.io}
\affiliation{%
  \institution{IoTeX}
  \city{Menlo Park}
  \country{US}
}

\author{Taeweon Suh}
\email{suhtw@korea.ac.kr}
\affiliation{%
  \institution{Korea University}
  \city{Seoul}
  \country{Korea}
}

\author{Weidong Shi}
\email{wshi3@central.uh.edu}
\affiliation{%
  \institution{University of Houston}
  \city{Houston}
  \country{US}
}

\author{Lei Xu}
\orcid{0000-0002-7662-2119}
\email{xuleimath@gmail.com}
\affiliation{%
  \institution{Kent State University}
  \city{Kent}
  \country{US}
}

% making it related to critical infrastructure
\begin{abstract}
    Trusted execution environment (TEE) technology has found many applications in mitigating various security risks in an efficient manner, which is attractive for critical infrastructure protection.
    First, the natural of critical infrastructure requires it to be well protected from various cyber attacks.
    Second, performance is usually important for critical infrastructure and it cannot afford an expensive protection mechanism.
    While a large number of TEE-based critical infrastructure protection systems have been proposed to address various security challenges (e.g., secure sensing and reliable control), most existing works ignore one important feature, i.e., devices comprised the critical infrastructure may be equipped with multiple incompatible TEE technologies and belongs to different owners.
    This feature makes it hard for these devices to establish mutual trust and form a unified TEE environment.
    To address these challenges and fully unleash the potential of TEE technology for critical infrastructure protection, we propose \sln{}, a decentralized coordination mechanism. 
    \sln{} uses blockchain technology to support key TEE functions in a heterogeneous TEE environment, especially the attestation service. 
    A Device equipped with one TEE can interact securely with the blockchain to verify whether another potential collaborating device claiming to have a different TEE meets the security requirements. 
    \sln{} is also flexible and can support new TEE schemes without affecting devices using existing TEEs that have been supported by the system.
\end{abstract}



\maketitle

%===============================
\section{Introduction}
Most modern critical infrastructures heavily rely on ICTs (information and communication technologies) to support their daily operations and provide uninterrupted services~\cite{xu2021reshaping,laplante2021artificial}.
The underlying ICT system can greatly improve the efficiency of a critical infrastructure, e.g., collecting information using IoT devices to monitor the status of major components and automating the operation in a precise manner.
While enjoying all the benefits, the deep integration with ICT also brings new security challenges to the critical infrastructures, e.g., valuable data collected to support the infrastructure can be leaked~\cite{jiang2020data}, the control software can be compromised~\cite{alladi2020industrial}, and many other cyber-based attacks that can disrupt the normal operation~\cite{thakur2016impact}.

Various techniques have been developed to mitigate these security/privacy concerns and protect the critical infrastructure from cyber attacks. 
For instance, communication encryption, malware detection and prevention, intrusion detection, and many other classical IT system protection mechanisms are used to enhance the robustness of critical infrastructure against cyber attacks.
Most of these methods focus on specific risks and try to fix known security issues.
While these methods have been proved to be effective for the designed purposes, they are not general and usually cannot handle new attacks.

The trusted execution environment (TEE) technology provides another option to improve the cybersecurity posture of critical infrastructures.
TEE relies on specific hardware to reduce the attack surface and prevent both existing and future risks, and provides a favorable balance among performance, functionality, and security assumptions. 
Specifically, a device equipped with TEE is capable of offering security protection for a wide range of applications. and its security primarily depends on the trustworthiness of a small piece of hardware (usually the CPU) and a secret key stored with the hardware\footnote{Here we do not consider risks such as side-channel attacks and hardware vulnerability.}.
Several works have been done on leveraging TEE for critical infrastructure protection~\cite{mcginthy2019secure,sebastian2019tee,coppolino2021protection}.

Currently, all major processor vendors have their own TEE solutions, including Intel SGX~\cite{intel2014sgx}, AMD SEV~\cite{Kaplan2016amd}, ARM TrustZone~\cite{ARM2009}, and Nvidia H100 GPU TEE~\cite{elster2022nvidia}.
Multiple TEE solutions are also developed for open-source RISC-V ISA~\cite{lu2021survey}.
Although these schemes follow the same design principles and have the same set of major components (i.e., isolation component and attestation component), they also have their own protocols and software stacks, which make it hard for them to interact with each others directly.
For example, if an IoT device equipped with TrustZone needs to send collected data to another component of the critical infrastructure for processing, which has servers with multiple TEE solutions, it is hard to build an end-to-end protection environment using TEE technology because they cannot work with each other.
Even for devices using the same TEE technology, they may still not be able to work together.
The reason is that an organization can set up its attestation service with its own root-of-trust to support all its own devices with TEE capability~\cite{scarlata2018supporting}, but this attestation service cannot recognize devices belong to other organizations (e.g., using different root-of-trusts).

A modern critical infrastructure is usually very complex and can involve devices from multiple parties, the fragmentation of of the TEE ecosystem makes it hard to deploy a heterogeneous TEE system for  protection.
Although there are several efforts on standardizing TEE, it is very unlikely one of them will dominate the market in a short term of period.
A straightforward approach to enable heterogeneous TEE is to modify each device to add support for other devices' TEE systems (either a different TEE scheme or a different root-of-trust).
However, this approach is not scalable as each device needs to support all TEE schemes used in the critical infrastructure.
Furthermore, when a device with a new type of TEE system/root-of-trust joins, all existing devices need to be updated.

To overcome these limitations and further unleash the potential of TEE technology for critical infrastructure protection, we propose \sln{}, which utilizes blockchain as the coordination and storage backbone to support the inter-operation of devices with different TEE technology/root-of-trust.
Assume multiple parties own and manage all ICT devices used in a critical infrastructure, these parties run a group of nodes to  maintain a blockchain system together.
This blockchain works as a translator to convert TEE-related messages in different formats to a common form.
For each device, we add a new component to interact with the blockchain and interpret the new common TEE message formats.
When a new type of TEE scheme is introduced, only the blockchain system needs to be updated and will not affect existing devices.
\sln{} brings several benefits:
\begin{itemize}
    \item It offers a unified framework to support heterogeneous a TEE system, which is attractive for complicated critical infrastructures protection;
    \item It reduces the system management complexity and can be easily extended to support new TEE schemes;
    \item It eliminates the single point of failure in the attestation process; and
    \item It protects information stored on the blockchain without affecting the function to support heterogeneous TEEs.
\end{itemize}
In summary, our contributions in the paper include:
\begin{itemize}
    \item We clarify the essential requirements on the design of a heterogeneous TEE system, especially for the scenario of critical infrastructure protection;
    \item We propose a detailed design of \sln{} that leverages blockchain technology to support the collaboration of multiple TEE schemes with multiple root-of-trusts; and
    \item Detailed analysis and evaluations are done to demonstrate the security and practicability of \sln{}.
\end{itemize}

The rest of the paper is organized as follows:
Section~\ref{sec-background} gives a quick review of the background of TEE.
Section~\ref{sec-overview} describes the high-level architecture of \sln{} and the detailed design of \sln{} is provided in Section~\ref{sec-detailed-design}.
Analysis and evaluation of \sln{} are presented in Section~\ref{sec-eval}.
Section~\ref{sec-related} reviews related works. 
We conclude the paper and discuss future works in Section~\ref{sec-conclusion}.

%===============================
\section{Background of TEE}\label{sec-background}
Hardware-based trusted execution environment (TEE) provides an effective and general mechanism to secure computation tasks.
In the remainder of the paper, when we use the term device, it means a computation device equipped with a TEE mechanism.
A TEE usually offers two basic functions: \textit{isolation} and \textit{attestation}.
\begin{itemize}
    \item \textit{Isolation}. An isolation mechanism creates a separate environment for a program to run and prevents other programs from intercepting or affecting its execution. 
    Isolation is a local protection mechanism and its implementation detail does not affect other devices collaborating with it.
    Therefore, the difference in isolation mechanisms is not a barrier to heterogeneous TEE construction.
    Note that isolation itself is not enough to build a TEE, because a remote user cannot verify whether it is interacting with a legitimate device equipped with TEE before sending its application to the device.
    \item \textit{Attestation}. The attestation function allows one party to verify the execution environment of a remote device, i.e., whether the device is equipped with legitimate hardware and running expected software. The attestation process is usually coupled with a key exchange protocol. If the attestation succeeds, both parties also share a secret, which is used to protect future communication. Because TEE provides an isolation mechanism, involved parties do not need to worry about the leakage of the shared secret key.
\end{itemize}

Another key concept related to TEE technology is \textit{root-of-trust}, which can also cause issues for inter-operations.
The root-of-trust is an entity who is authorized to issue identities to TEE hardware, and these identities will be used to support other TEE related operations, especially the attestation.
In practice, the root-of-trust can be a public/private key pair managed by the vendor/owner, which is similar to the public/private key owned by a certificate authority (CA) of a PKI, i.e., the private key is used to issue certificates for TEEs, and the public key is distributed for others to verify the certificate.

\begin{figure}
    \centering
    \includegraphics[width=3.4in]{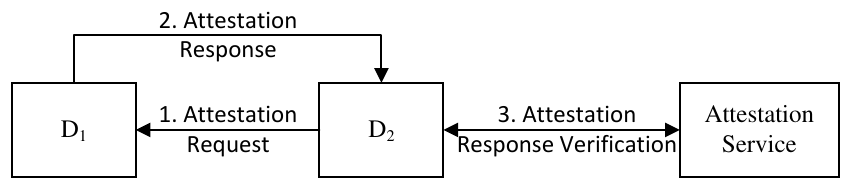}
    \caption{The abstract model of attestation protocol. $D_2$ utilizes the attestation protocol to verify $D_1$. 
    If attestation succeeds, both devices also obtain a shared secret key through a key agreement protocol nested within the attestation process.
    They can use the shared secret key to exchange information safely after the attestation.}
    \label{fig-attestation-model}
\end{figure}

%\subsection{Abstract Attestation Model}
Different hardware vendors usually have their own attestation protocols, and even the same vendor may support multiple attestation protocols for their products. 
These attestation protocols share a similar abstract model, which is summarized in \figurename~\ref{fig-attestation-model}.
Two devices \textit{Device 1} (denoted as $D_1$) and \textit{Device 2} (denoted as $D_2$) belong to the same owner (i.e., sharing the same root-of-trust) and support the same attestation protocol and both register with the \textit{Attestation Service} ($AS$) before they can verify each other.
Without loss of generality, \figurename~\ref{fig-attestation-model} only considers one-way attestation, i.e., $D_2$ verifies whether $D_1$ has a claimed TEE running.
They can run the same protocol in the other direction to achieve mutual attestation.
Each device has a public/private key pair $(pk_{D_1}, sk_{D_1})$, where the private key is kept inside the device in a secure manner, and the public key is registered to the attestation service.
A device also keeps a copy of the public key of the attestation service to verify the authenticity of messages the service creates.%(denoted as $pk_{AS}$).

To verify the execution environment of $D_1$, a pre-configured program in $D_1$ measures the target environment creates an attestation report of the environment, and signs the report with the private key $sk_{D_1}$.
Both the report and signature are then sent to $D_2$, but $D_2$ cannot verify the signature directly as it does not manage all registered devices and corresponding public keys.
Instead, $D_2$ forwards received information to $AS$ to process.
$AS$ uses previously registered public key $pk_{D_1}$ to verify whether the received report and signature are consistent and notifies $D_2$ of the verification result.
$D_2$ has a copy of $AS$'s public key, so it can verify the authenticity of the received result.

%==================================
\section{Overview of \sln{}}\label{sec-overview}
This section provides an overview of \sln{} and outlines its design goals.

\subsection{Major Participants}
\begin{figure}
    \centering
    \includegraphics[width=3.2in]{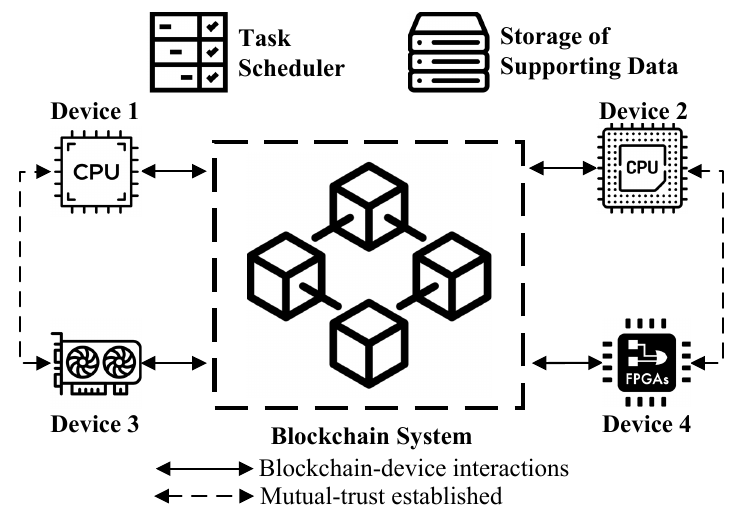}
    \caption{High level architecture of \sln{}. 
    Each computation device has its own TEE hardware and protocols. These TEE schemes are not compatible with each other for different reasons. 
    When two devices need to collaborate to finish a computation task, they establish mutual trust with the help of the blockchain system and TEE schemes, i.e., the devices verify each other's execution environment even if they use different TEE schemes, and establish a shared secret key for information exchange. 
    After mutual trust is established, the two devices can interact directly without relying on the blockchain.}
    \label{fig-system-overview}
\end{figure}

Here we focus on those components that are related to TEE but ignore others involved in a critical infrastructure.
\figurename~\ref{fig-system-overview} provides an overview of the abstract structure of \sln{}, and there are mainly two types of participants involved:
\begin{itemize}
    \item \textit{Devices}. 
        The system involves a large number of computation devices, and each of them is equipped with a TEE.  
        Multiple devices need to work together to finish a computing task in a secure manner, i.e., the computation results produced by TEEs of different devices are exchanged securely but the information will not be leaked to a non-TEE device unless it is the task's requirement. 
        These devices use different TEE solutions, and they are owned and managed by different entities.%, e.g., an adversary device may pretend to have a valid TEE to obtain information from other devices.
    \item \textit{Blockchain}. 
        The other important component is the blockchain system, which works as the coordinator between the devices and offers attestation services. 
        Nodes of the blockchain run a consensus protocol to make various decisions on the system's operation. 
        Although some nodes of the blockchain can be compromised, a device always trusts the decisions made through the consensus protocol and stored on the ledger. 
        The nature of our application requires that only authorized nodes are allowed to maintain the blockchain. Therefore, the blockchain is permissioned and each node has an identity (in the form of a public/private key pair).
\end{itemize}

Besides the blockchain and devices, there are two other participants (as shown on top of \figurename~\ref{fig-system-overview}), the \textit{task scheduler} and the \textit{storage of supporting data}.
\begin{itemize}
    \item The \textit{task scheduler} helps allocate adequate devices for computation tasks based on the requirements of the tasks and the device-supported features. The task scheduler is not trusted and may schedule a malicious device for tasks. 
    Note that the task scheduler is not an essential part of \sln{}, and its functions can also be implemented as part of the blockchain and/or the devices.
    For instance, when a device needs to collaborate with another device, it sends a request directly to that device if it has its identity.
    In case the device only knows the requirements but does not have the address/identity of the target device, it relies on the task scheduler to select one.
    In the context of critical infrastructure, it is also possible that the task assignments are pre-determined, and this participant is not necessary.
    \item The \textit{storage of supporting data} manages information that is necessary for \sln{} operations. In practice, it can be implemented as either a centralized or a decentralized manner. Integrity tags (e.g., hash values) of supporting data can be stored on the blockchain to prevent unauthorized modification of the key data. The concrete information managed by this participant is discussed in later sections.
\end{itemize}

\subsection{Assumptions and Design Goals of \sln{}}\label{subsec-assumption-goal}

\para{Design goals of \sln{}}
\sln{} is designed to support the inter-operation of heterogeneous TEE schemes, and devices of the same TEE but managed by different entities.
Concretely, the design goals of \sln{} include:
\begin{itemize}
    \item Multiple devices with different TEEs can collaborate to provide a unified trusted execution environment to finish a computation task;
    \item Devices with new TEEs can be added to \sln{} easily to work with existing devices, and existing devices do not need to be modified; and
    \item The unified TEE created by \sln{} can offer all security features supported by a typical TEE scheme.
\end{itemize}

\para{Technical challenges of \sln{}}
There are two major challenges for \sln{} to support a heterogeneous TEE system and achieve the design goals, both of which are related to the remote attestation process.
\begin{itemize}
    \item Compatibility of the root of trust. A TEE is identified by a public/private key pair, which is issued and endorsed by a corresponding root-of-trust (RoT). However, two identities issued by different RoTs for two TEEs cannot be recognized by the other device. Therefore, they cannot establish mutual trust.
    \item Compatibility of data formats. A TEE solution has its own format to describe the execution environment (e.g., software stack and hardware information), which is used for attestation reports. Two devices equipped with different TEEs cannot understand each other's attestation report.
    %\sln{} defines a set of A devices in a heterogeneous system that may not understand data records created by another device.
\end{itemize}

\para{Assumptions}
The design of \sln{} relies on several assumptions, which are summarized as follows:
%\sln{} is designed to support the interoperation of devices with heterogeneous TEEs, we need to make some basic assumptions, 
\begin{itemize}
    \item \textit{Permissioned blockchain.} The blockchain stores various device information and only permissioned blockchain is considered in the design of \sln{}.
    
    \item \textit{Attribute equivalence.} The attestation report of a TEE consists of a sequence of software/hardware attributes. Although a TEE scheme uses its own language to describe these attributes, we assume there exist a set of equivalent relationships between attributes of different TEE schemes. For instance, two attributes that represent the same software on different hardware platforms are equivalent.
    
    \item \textit{Common cryptographic suites.} All participating devices support certain cryptographic tools, including symmetric encryption (e.g., AES with a secure operation mode) and key exchange (e.g., Diffie-Hellman key exchange). This assumption is necessary as two participants need to exchange information directly in a secure manner after mutual trust is established. 
    
    \item \textit{Simplified TEE key architecture.} A TEE scheme can have a complex key derivation architecture, i.e., multiple public/private key pairs are derived from various sources of the device and used for different purposes. The concrete key derivation mechanism does not affect the operation of \sln{} and for simplicity, we assume a device uses a single key pair for various purposes, including identification and attestation endorsement.
\end{itemize}

%==================================
\section{Detailed Design of \sln{}}\label{sec-detailed-design}
This section presents the detailed design of \sln{}.
Without loss of generality, we consider the situation that two heterogeneous TEEs collaborate with the help of \sln{}, and the scheme can be easily extended to the case of an arbitrary number of devices and types.

\subsection{Supporting Data Structures}
\sln{} utilizes the following key data structures to support a heterogeneous TEE system.
\begin{itemize}
    \item \textit{Device registration information.}
    As discussed in Section~\ref{sec-background}, a device has a public/private key pair assigned to its TEE, where the private key is kept securely inside the device TEE and the public key is registered to the system.
    Most existing TEE solutions store public key information in a centralized system.
    For \sln{}, this information is stored on the blockchain, i.e., each node of the blockchain keeps a copy of the public keys of all registered devices. Detailed registration/enrollment protocol is described in Section~\ref{subsec-enrollment}.
    
    \item \textit{Environment attribute description and equivalent relationship.}
    An attestation report includes information on the hardware and software that is involved in the construction of an isolated execution environment.
    \sln{} maintains a database containing two types of records:
    \begin{itemize}
        \item Records of attributes. This type of record save software/hardware attributes of all supported TEE schemes.
        \item Records of attribute equivalence. Based on the assumptions given in Section~\ref{subsec-assumption-goal}, some attributes from different TEE schemes are equivalent. This type of record stores such equivalence information.
    \end{itemize}
    In practice, these two types of records can be kept in the same data structure. 
    For instance, for each software/hardware attribute, \sln{} creates a tuple, which includes a unique identity number, and a list of equivalent software/hardware attributes.
    The tuples can either be stored on-chain or off-chain, as long as all blockchain nodes reach a consensus on existing tuples.
    
    \item \textit{Common attestation report verification result.}
    According to the design of \sln{}, all participating devices go to the blockchain-based attestation service to obtain attestation report verification results.
    \sln{} defines a common format for the verification result, and all devices in the system need to support such format.
\end{itemize}

\subsection{Attestation Protocol of \sln{}}
Attestation is the fundamental component that supports most other operations of \sln{}.
The attestation scheme of \sln{} consists of the following sub-protocols:
\begin{itemize}
    \item \attrqst. A device uses the \attrqst{} algorithm to create a request to verify the execution environment of another device, which it needs to collaborate with to finish a computation task.
    \item \attgen. A device uses the \attgen{} algorithm to create an attestation report and submit it to the blockchain-based attestation service in response to a request created by \attrqst{}.
    \item \attconvrfy. The blockchain nodes run the \attconvrfy{} algorithm together to first verify the validity of a received attestation report created by \attgen{} and convert the result to the common attestation report format.
    \item \attvrfy. A device uses the \attvrfy{} algorithm to interact with the blockchain system to obtain and process the verification result stored on the blockchain.
\end{itemize}

In the following, we describe the concrete construction of each sub-protocol.
Without loss of generality, we assume device $D_2$ wants to verify the execution environment of device $D_1$, and they support different TEE schemes.

\para{\attrqst}
$D_2$ runs \attrqst{} to create the request to verify the execution environment of another device $D_1$, which is in the following form:
\begin{equation}\label{eq-e1-rqst}
    \rqst_{D_1} \leftarrow (\mathit{lst}, \mathit{id}_{D_2}, \sig{D_2}{\rqst{}}).
\end{equation}
Here $\mathit{lst}$ is a list of software/hardware attributes that $D_2$ needs from another device to finish the computation task,
$\mathit{id}_{D_2}$ is the identity of $D_2$, 
and $\sigma_{D_2}^\rqst{}$ is the digital signature of $D_2$ on the request.

The request $\rqst_{D_1}$ needs to be sent to the corresponding device, which can then respond to the request.
There are two cases:
\begin{itemize}
    \item \textit{Case 1}. If $D_2$ knows the identity of the device it wants to collaborate, e.g., a device with identity $\mathit{id}_{D_1}$, then $\mathit{id}_{D_1}$ is included in the attributes list $\mathit{lst}$.
    After the transaction is accepted by the blockchain, the corresponding request is either forwarded to the device $D_1$ by a blockchain node or retrieved by the device $D_1$ directly.
    \item \textit{Case 2}. For certain scenarios, e.g., the cloud environment, $D_2$ may not know the identity of the device that it needs to work with, but only has the requirements for the device (described in $\mathit{lst}$).
    In this case, a dedicated scheduler (i.e., \textit{task scheduler} in \figurename~\ref{fig-system-overview}) is responsible for finding a device that meets the requirements of the given request.
\end{itemize}

For \textit{Case 2}, it is possible that the task scheduler is malicious according to the security assumption, and it can select the wrong device for $D_2$ to work with.
This is not a concern as \sln{} will help $D_2$ to verify the response in a later stage.

\para{\attgen}
Without loss of generality, we assume the device $D_1$ receives the request $\rqst_{D_1}$ created in the previous sub-protocol.
$D_1$ runs \attgen{} to create an attestation report of its local environment for the request from $D_2$, which is given in \eqname~\eqref{eq-native-att-report}.

With the request ($\rqst_{D_1}$ from $D_2$), \attgen{} calls the native attestation report generation mechanism of $D_1$ to generate an attestation of the created environment, which is denoted as:
\begin{equation}\label{eq-native-att-report}
    (\rpt_\natv, k_{D_1}, \mathit{id}_{D_2}, \sigma_{D_1}),
\end{equation}
In \eqname~\eqref{eq-native-att-report},
$\rpt_\natv$ is the attestation report consisting of multiple attributes, and each attribute is registered to \sln{};
$k_{D_1}$ is the value selected by $D_1$ for key agreement; 
$\mathit{id}_{D_2}$ is the identity of the requester; and
$\sigma_{D_1}$ is the signature of the previous parts of the message generated with $D_1$'s private key for attestation report authentication.
The created attestation report is sent to the blockchain component of \sln{} for further processing.

\para{\attconvrfy}
A blockchain node $bn$ runs \attconvrfy{} to process the received message described in \eqname~\eqref{eq-native-att-report}.
Since the device $D_1$ that creates the attestation message is registered to \sln{}, $bn$ has a copy of $D_1$'s public key $pk_{D_1}$ for attestation verification.
$bn$ verifies two facts:
\begin{itemize}%[(i)]
    \item The attestation report is valid. This is done by verifying the attached digital signature of the attestation report; and
    \item The environment provided by $D_1$ (described in the attestation report) meets the requirements of $D_2$ (described in the attributes list). This is done by comparing $\mathit{lst}$ with $\rpt_\natv$. Note that the blockchain nodes have the knowledge of attributes equivalence in different environments, and $bn$ can decide whether $\rpt_\natv$ meets the requirements given in $\mathit{lst}$ even if they are for two different types of devices.
\end{itemize}

The blockchain node $bn$ creates a transaction on the verification result.
Similar to other blockchain systems, $bn$ broadcasts the transaction to other nodes of the blockchain.
All nodes of the blockchain run the consensus protocol to determine whether to include the transaction in the blockchain.
Unlike most existing TEE schemes, the blockchain does not actively notify $D_2$ on the verification result, but only stores the transformed verification information on the blockchain.
%In practice, \sln{} can either require $D_2$ to monitor the blockchain for the verification result or use the task scheduler to notify $D_2$.

\para{\attvrfy}
As described in Section~\ref{sec-overview}, $D_2$ does not know the TEE scheme of $D_1$, but trusts the result accepted on the blockchain.
$D_2$ runs the \attvrfy{} sub-protocol to interact with the blockchain to obtain the verification result of the execution environment provided by $D_1$, which consists of the following major steps:
\begin{itemize}
    
    \item Obtaining and verifying the corresponding verification result. After \attvrfy{} detects the related transaction is posted on the blockchain, it obtains a copy of the transaction from the blockchain. Based on the security assumption, $D_2$ trusts the result stored on the blockchain, i.e., whether the environment provided by $D_1$ meets its requirements. Detailed design of interacting with the blockchain is provided in Section~\ref{subsec-blockchain-interaction}.
    \item Establishing a secure connection. If the verification result is positive, $D_2$ can start to build a secure connection with $D_1$, which can be done in different ways. For instance, running the attestation protocol in the other direction to pass a key agreement input to $D_1$. 

\end{itemize}

\subsection{Device Enrollment}\label{subsec-enrollment}
In the attestation protocol of \sln{}, we assume participating devices are enrolled in \sln{} and each blockchain node has a copy of its public key used for attestation report verification.
There are two ways to enroll a device:
\begin{itemize}
    \item The first way of enrollment relies on the device vendor. Some TEE schemes support third-party attestation services, such as the Intel DCAP~\cite{sardar2020formal}. For such a TEE scheme, the vendor works as the root-of-trust (i.e., it contributes to the device TEE root key pair generation and maintains a copy of the public key) and shares the public key with the third party. In \sln{}, the vendor signs the device attestation public key and shares both messages with blockchain nodes. A blockchain node keeps a copy of the vendor's public key and verifies the signature of the device key before accepting it.
    
    \item The second way of enrollment requires a blockchain node to interact with the device directly to obtain the key material without going through the network. Xu \emph{et. al.} ~\cite{xu2020fpga} proposed to pass a device to multiple parties physically to allow them to build the attestation key pair in a collaborative/decentralized way. The same idea can be applied to help the blockchain nodes to obtain a device's attestation public key. By physically connecting to the device, the public key is passed to the node without relying on a third party to authenticate it.
\end{itemize}
The first approach requires the TEE scheme to support third-party attestation.
The second approach supports the construction of a decentralized root of trust, but requires the device hardware to offer an interface for attestation public key export and is more expensive in practice.
The enrollment mechanism is relatively independent of other parts of \sln{}, and \sln{} can support both device enrollment methods to allow interactions between different types of TEE schemes.

\subsection{Device-Blockchain Interactions}\label{subsec-blockchain-interaction}
The blockchain works as the bridge between $D_1$ and $D_2$, and both of them need to interact with the blockchain in order to establish mutual trust.
The blockchain is a decentralized system, and it is only trusted as a whole, i.e., some nodes may not follow the pre-defined protocols and provide wrong information. 
The design of the underlying blockchain, especially the consensus protocol, affects the way to protect the interactions between a device and the blockchain.

\para{Sending transactions to the blockchain}
It is relatively easy for a device to send a transaction (e.g., attestation request and attestation report) to the blockchain.
The device keeps a list of blockchain node addresses. 
When it generates a transaction, it either multicasts it to all nodes on the list or tries to send it to them sequentially if a node does not respond within a time frame.
As long as the blockchain as a whole is trusted, the device can reach some honest nodes, and the transaction is processed correctly.
At the same time, invalid transactions are rejected by the blockchain consensus mechanism.

\para{Obtaining information from the blockchain}
A participating device is only interested in transactions that are related to it, and it is a waste of resources for the device to keep a copy of the whole blockchain.
Furthermore, a device can be an IoT device with limited storage capacity.
Therefore, a device does not participate in the maintenance of the blockchain. 
On the other hand, the device must be able to verify whether a transaction has been accepted by the blockchain because a malicious blockchain node may try to cheat the device by providing invalid transactions to it.
Depending on the underlying blockchain, \sln{} uses different methods to guarantee that the device obtains transactions from the blockchain in a reliable manner.
For instance, if the blockchain is built using proof-of-work and the longest-chain principle (e.g., Bitcoin), a device only accepts a block if there are more than a certain number of blocks have been added after it.
As \sln{} considers only permissioned blockchain, more efficient blockchain schemes can be utilized, such as those based on BFT and proof-of-authority (PoA).
If a BFT-based blockchain (e.g., \cite{luu2015scp}) or a PoA-based blockchain (e.g., \cite{szilagyi2017eip}) is used, the device can interact with blockchain nodes participating in the BFT or the authority nodes to determine whether a block has been accepted by the blockchain.

\subsection{Adding Support of New TEE Scheme}
A naive approach to enable inter-operation between two different TEE schemes is to add an extra support component to one of them.
This new component can work as a proxy to translate inbound/outbound messages to enable the collaboration.
The major issue with this approach is scalability. 
Every time a new type of TEE scheme is introduced, all existing devices have to update to support the new scheme.

\sln{} addresses this issue by shifting all translation jobs to the blockchain. 
To add a new TEE scheme to \sln{}, we do not need to modify information/configuration of any existing TEE schemes.
\sln{} uses the following steps to add support for a new TEE scheme:
\begin{enumerate}
    \item Adding the attributes (e.g., information about the software stack and cryptographic algorithms suit) for the new environment description, and establishing their equivalence information with existing attributes of other TEEs;
    \item Adding necessary algorithms (e.g., digital signature verification) that are needed to verify the attestation report of the new TEE scheme.
\end{enumerate}
After these two steps are done, devices with the new TEE scheme can register to \sln{} and interact with existing devices.

\subsection{\sln{} Client}
To support interactions with the \sln{} blockchain and other trusted execution environments, each device participating \sln{} runs a small client in its local TEE together with the concrete application.
This client is mainly responsible for interacting with the blockchain, including:
\begin{itemize}
    \item Interacting with the blockchain. The client prepares/submits transactions to the blockchain, and retrieves/verifies transactions from the blockchain;
    \item Processing transactions. From a security perspective, the processing of transactions mainly consists of two tasks, i.e., verification of digital signatures (verifying attestation results provided by the blockchain) and running key agreement protocols (establishing secure communication channels with remote devices).
\end{itemize}

%==================================
%\section{Cross Device Communication}\label{sec-impl}
\section{A Use Case of \sln{}}\label{sec-impl}
%This section looks at a use case for remote attestation and cross-device communication between an Intel SGX device and an AMD SEV-verified device.
This section presents the detailed interactions between two machines using different TEE schemes to leverage \sln{} to establish mutual trust.
Without loss of generality, we consider a machine with Intel SGX (denoted as $D_2$) and a machine with AMD SEV (denoted as $D_1$), and the Intel SGX machine $D_2$ initializes the process.
We also assume $D_2$ knows the identity of $D_1$ in advance.

\sln{} relies on existing TEE APIs to finish the task, and we summarize the major APIs of both platforms in \tablename~\ref{tab-apis}.

\begin{table}
\caption{The major APIs of Intel SGX and AMD SEV TEE.}\label{tab-apis}
\begin{tabularx}{\columnwidth}{p{0.5in}|p{0.7in}|p{0.7in}|p{0.8in}}
\textbf{Platform} & \textbf{Function}    &   \textbf{Instruction} & \textbf{Description}\\
\hline
Intel  & Enclave Initialization & \texttt{ECREATE}     & Create an enclave\\
       &                        & \texttt{EINIT}       & Initialize an enclave\\
       &                        & \texttt{EENTER}       & Enter an enclave \\
       & Generate Cryptographic Keys    & \texttt{EGETKEY}     & Create a cryptographic key\\
       &                                & \texttt{EREPORT}     & Create a cryptographic report\\
       && \texttt{EEXIT}        & Exit an enclave\\
\hline
AMD & Enclave Initialization & \texttt{INIT}          & Create an enclave\\
    & & \texttt{INIT\_EX}      & Initialize an enclave\\
    & & \texttt{ATTESTATON}    & Create a cryptographic key\\
    & & \texttt{EREPORT}       & Create a cryptographic report  \\
%    & Platform API for launching encrypted Guest & \texttt{LAUNCH \_START} & Begin to launch a new SEV enabled guest \\
%    & & \texttt{LAUNCH\_UPDATE\_DATA}   & Encrypt guest data for launch \\
%    && \texttt{LAUNCH\_FINISH} & Complete launch of guest\\
\end{tabularx}
\end{table}

The detailed interactions between $D_1$ and $D_2$ are as follows:
\begin{enumerate}
    \item The Intel SGX machine $D_2$ runs \attrqst{} to create the request to verify the execution environment of the AMD device $D_1$. \attrqst{} is implemented with the following  API calls and operations:
    \begin{itemize}
        \item $D_2$ calls \texttt{ECREAT} and \texttt{EINIT} to create a local trusted execution environment to prepare for later interactions;
        \item $D_2$ loads \sln{} client code into the enclave, which is used to create the request (the format is described in \eqname~\eqref{eq-e1-rqst}) to verify an execution environment of $D_1$;
        \item The created request is sent to the blockchain of \sln{}, and the blockchain forwards the request to $D_1$.
    \end{itemize}
    \item The AMD SEV machine $D_1$ runs \attgen{} to create an attestation report. \attgen{} is implemented with the following API calls and operations:
    \begin{itemize}
        \item If $D_1$ does not have an active trusted execution environment running the \sln{} client, it runs \texttt{INIT} and \texttt{INIT\_EX} to create one and loads the client;
        \item $D_1$ then runs the attestation generation API \texttt{EREPORT} to create a report (the attestation report also includes key agreement information);  
        \item The attestation report is then returned to the blockchain for further processing.
    \end{itemize}
    \item The \sln{} blockchain runs \attconvrfy{} to process the attestation report received from $D_1$. Note that AMD SEV does not have a dedicated API for attestation report verification, and \attconvrfy{} is implemented with the following API calls and operations:
    \begin{itemize}
        \item The blockchain node contacts the AMD key server to obtain necessary certificates. In the current SEV design, the AMD Key Server provides means to retrieve a CPU-specific CEK certificate for a given platform ID~\cite{SEV_CEK_Server}, which serves as the root certificate.
        \item The blockchain node verifies the certificates chain and extracts the public key for $D_1$ in a reliable way, and uses it to verify the digital signature of the attestation report received from $D_1$;
        \item If all verifications succeed, the blockchain node converts the attestation report (together with key agreement information) to the \sln{} format that can be consumed by $D_2$ (through the \sln{} client).
    \end{itemize}
    \item The Intel SGX machine $D_2$ runs \attvrfy{} to finish the attestation of $D_1$. This is done by the \sln{} client loaded into its own trusted execution environment.
\end{enumerate}

\begin{figure*}
    \centering
    \includegraphics[scale=0.35]{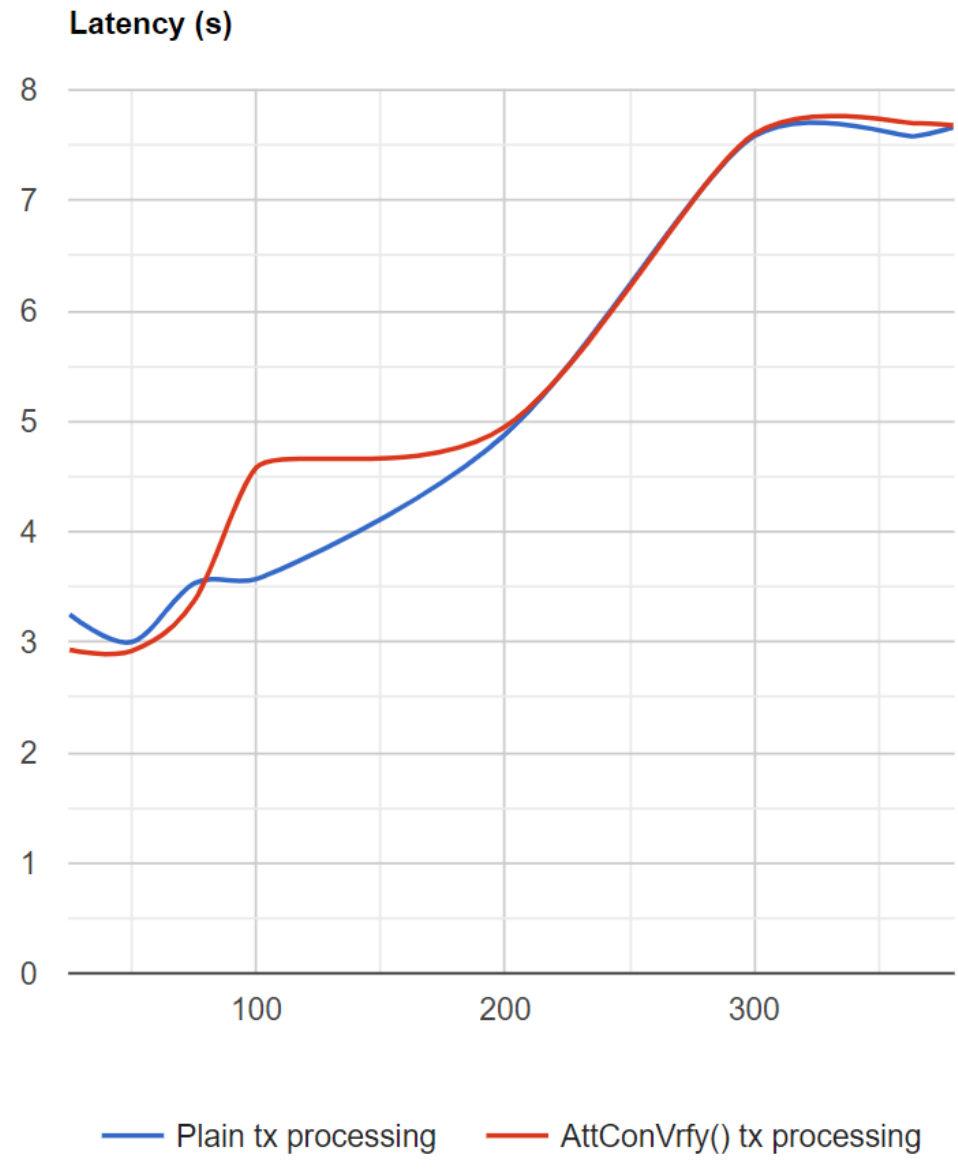}
    \hspace{0.4in}
    \includegraphics[scale=0.35]{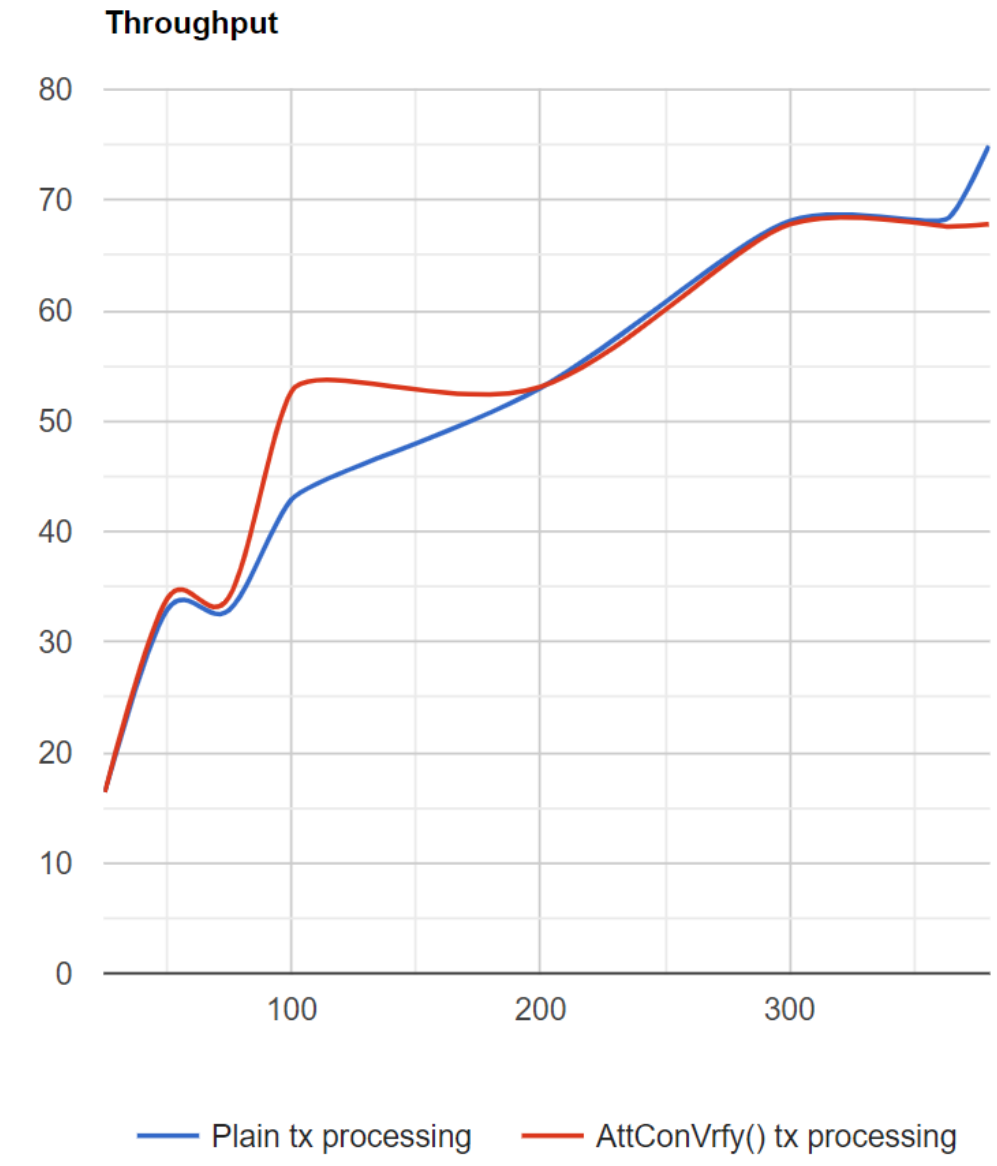}
    \caption{Blockchain related performance. The x-axis is the transaction submission rate (transactions/second), and the y-axis is the corresponding performance metric.}
    \label{fig-perf}
\end{figure*}

%==================================
\section{Analysis and Evaluation}\label{sec-eval}
This section analyzes the security features of \sln{} and evaluates its performance.

\subsection{Security Analysis}
Compared with existing TEE schemes that work for the same type of devices, \sln{} mainly modifies three places in the TEE design: enrollment, the attestation report verification, and verification of the verification result.
Assuming all existing TEE schemes involved in \sln{} are secure, we only need to argue that these modifications do not introduce new vulnerabilities.

\para{Security of the enrollment}
Device enrollment is the process of registering the public key for attestation to each blockchain node, and \sln{} supports two device enrollment methods.
\begin{itemize}
    \item Obtaining public key from the vendor. Assuming the device vendor is honest and supports third-party attestation service, an honest blockchain node can obtain the attestation public key correctly, as this protocol is equivalent to setting up a third-party attestation service.
    \item Physically obtaining the device attestation public key. As long as the device hardware is correctly implemented (e.g., without trojan and backdoor), the blockchain node can safely extract the public key from the device directly through a physical connection. Unless an adversary can control this physical connection, it is hard for the attacker compromising this process.
\end{itemize}

\para{Security of attestation report verification} 
The attestation report created by a device is submitted to the blockchain for verification. As each blockchain node has the public key associated with the digital signature on the attestation report, it can verify the validity of the report by checking the associated digital signature. The blockchain as a whole is trusted, and the consensus mechanism guarantees that only the correct attestation report verification result is included in the blockchain. 

\para{Security of the verification result}
For a classical TEE scheme, the device obtains the result from the attestation service and verifies its digital signature on the result. With \sln{}, the major difference is that the device fetches the result directly from the blockchain, which is a distributed system. With the method described in Section~\ref{subsec-blockchain-interaction}, \sln{} guarantees that a transaction the device obtained has been included in the blockchain.

In summary, \sln{} achieves the design goals of supporting heterogeneous TEE without sacrificing security.
Furthermore, as \sln{} utilizes a blockchain with multiple nodes in the verification of an attestation report, it avoids the potential single point of failure of the attestation service and is more robust even if the system only has devices of the same type of TEE.

\subsection{Performance and Cost Evaluation}
The performance and cost of \sln{} are considered at two levels, \textit{device level} and \textit{system level}.
At the device level, the extra cost is mainly related to obtaining attestation verification results from the blockchain, which usually involves a couple of digital signature verifications depending on the underlying blockchain.
At the system level, the heaviest part is the blockchain operations, especially \attconvrfy{}, which requires the blockchain to verify an incoming attestation report and store the converted result.

To evaluate the performance, Hyperledger Besu (Ethereum)~\cite{samuel2021choice} with four-nodes IBFT consensus protocol~\cite{saltini2019correctness} is used.
The hardware platform for the evaluation is a Ubuntu server with an Intel Xeon CPU E5-2620 v4 (8 Cores, 16 Threads) and 128 GB memory.
We conduct two experiments. 
In the first experiment, we submit plain transactions to the blockchain.
In the second experiment, \attconvrfy{} is added as part of the transaction processing procedure.
Experiment results are summarized in \figurename~\ref{fig-perf}, where the left figure shows latency data and the right one shows throughput data.
It can be seen that adding \attvrfy{} does not change the performance of the underlying blockchain system significantly.

%==================================
\section{Related Works}\label{sec-related}
This section reviews works that are related to the design of \sln{}.

\para{Blockchain and TEE}
There are mainly two research directions on the integration of blockchain and TEE.
The first direction is applying TEE technology for blockchain and its application.
Many works are done on using TEE for efficient blockchain construction~\cite{wang2020hybridchain,fu2022teegraph}.
As a hardware-based isolation mechanism, TEE is also used to improve the security and privacy of blockchain-based applications~\cite{yuan2018shadoweth,yan2020confidentiality,karanjai2022lessons,cheng2019ekiden}.
The second direction is applying blockchain technology to improve TEE.
Xu \emph{et.al.} proposed to use blockchain to build decentralized root-of-trust~\cite{xu2020fpga}.
Neither direction considers the problem of supporting heterogeneous TEE schemes.

\para{Protection of heterogeneous computation environment}
To tackle challenging computation tasks, it is attractive to use multiple types of computation devices.
When the task is sensitive, it is a natural idea to use TEE to protect it from both external and internal adversaries.
Most existing works mainly rely on the TEE of a single type of device to carry out the computation in a secure manner.
For instance, HETEE divides the task into sensitive and non-sensitive parts. 
Sensitive sub-tasks are executed on CPUs with TEE and non-sensitive sub-tasks are scheduled to GPUs in the same rack to improve performance~\cite{zhu2020enabling}.
There are also works that consider supporting a fixed heterogeneous TEE system.
Xia \emph{et.al.} proposed a heterogeneous TEE of Intel SGX and FPGA~\cite{xia2021sgx}, and ARM research developed Veracruz framework ~\cite{arm2021veracruz} that supports Intel SGX, ARM TrustZone, ARM CCA and AWS Nitro.
Compared with these works, \sln{} has the capability to support an arbitrary set of TEE schemes and also considers the collaboration of multiple entities who manage their own devices.

\para{TEE for critical infrastructure protection}
TEE finds a wide range of applications to enhance the security of different types of critical infrastructures, including trusted monitoring~\cite{jung2022trusted}, secure node construction~\cite{mcginthy2019secure}.
There are also works that focus on applying TEE technologies to improve critical infrastructure for specific industries, like power grid management~\cite{sebastian2019tee} and chemical storage~\cite{coppolino2021protection}.
Most of these works only consider the desirable security features TEE offers and their applications in critical infrastructure, but ignore the challenges of using of different types of TEE solutions in the same system.
Therefore, the decentralized ledger-based heterogeneous TEE system proposed in this paper is largely complementary 

%==================================
\section{Conclusion and Future Work}\label{sec-conclusion}
Providing a unified and transparent TEE that includes various computation devices is not only important to support various TEE-based protection mechanisms for complex critical infrastructures, but also useful for a large range of security-sensitive tasks that require collaborations of multiple types of devices. 
The design of \sln{} addresses two major obstacles in the construction of a heterogeneous TEE, the incompatibility of different TEE protocols and the lack of trustworthiness between different device owners (i.e., multiple root-of-trusts) by utilizing the emerging blockchain technology. 
Furthermore, \sln{} minimizes the modification of existing TEE schemes and only relies on a small set of assumptions.

For future works, we plan to conduct the following research and development tasks:
\begin{itemize}
    \item Implementing \sln{} client in more TEE systems. This work focuses on classical TEE systems and there are several more emerging types of TEE  will be considered, including FPGA and GPU. These two types of computation devices have been offered in the cloud environment and \sln{} can provide a unified TEE for a broad range of applications.
    \item Enhancing the privacy. Current design of \sln{} focuses on security and privacy is not considered, e.g., the blockchain can learn the interaction history between participating computation devices. Although some TEE systems  support privacy preserving attestation (e.g., Intel EPID~\cite{sardar2020towards}), it cannot be adopted directly by \sln{} as other TEE systems may not support it.
    \item Explore new applications. TEE has a wide range of potential applications, including secure communication, data storage, and smart contracts. In the future, it will be important to continue exploring new ways that these systems can be used to solve real-world problems and improve people's lives.
\end{itemize}

\begin{acks}
This research is partially supported by The Research Council of Kent State University.
\end{acks}

\balance
\bibliographystyle{ACM-Reference-Format}
\bibliography{tidy}

\end{document}